\documentclass[a4paper,12pt]{spieman}  
\usepackage{amsmath,amsfonts,amssymb}
\usepackage{graphicx}
\usepackage{setspace}
\usepackage{tocloft}

\title{Ray mapping approach in double freeform surface design for
collimated beam shaping beyond the paraxial approximation}

\author[a,*]{Christoph B\"osel}
\author[a]{Norman G. Worku}
\author[a,b]{Herbert Gross}
\affil[a]{Friedrich-Schiller-Universit\"at Jena, Institute of Applied Physics, Albert-Einstein-Str. 15, 07745 Jena, Germany}
\affil[b]{Fraunhofer-Institut f\"ur Angewandte Optik und Feinmechanik, Albert-Einstein-Str. 7, 07745 Jena, Germany}

\cftpagenumbersoff{figure}
\cftpagenumbersoff{table} 
\begin{document} 
\maketitle

\begin{abstract}
Numerous applications require the simultaneous redistribution of the irradiance and phase of a laser beam. The beam shape is thereby determined by the respective application. An elegant way to control the irradiance and phase at the same time is from double freeform surfaces.
In this work the numerical design of continuous double freeform surfaces from ray mapping methods for collimated beam shaping with arbitrary irradiances is considered. These methods consist of the calculation of a proper ray mapping between the source and the target irradiance and the subsequent construction of the freeform surfaces.
By combining the law of refraction, the constant optical path length and the surface continuity condition, a partial differential equation (PDE) for the ray mapping is derived. It is shown that the PDE can be fulfilled in a small-angle approximation by a mapping derived from optimal mass transport with a quadratic cost function.
To overcome the restriction to the paraxial regime we use this mapping as an initial iterate for the simultaneous solution of the Jacobian equation and the ray mapping PDE by an optimization. The presented approach enables the efficient calculation of compact double freeform surfaces for complex target irradiances. This is demonstrated by applying it to the design of a single-lens and a two-lens system.
\end{abstract}


{\noindent \footnotesize\textbf{*}\linkable{christoph.boesel@uni-jena.de} }

\begin{spacing}{2}   

\section{Introduction}
\label{sec:1} 

In recent years the manufacturing of freeform surfaces has become increasingly feasible. These freeform surfaces offer an elegant way of simultaneous irradiance and phase control. Therefore, the development of numerical algorithms for the calculation of continuous freeform surfaces for control of the irradiance and/or the phase of a beam is of great interest.
\\
In this work the problem of designing continuous double freeform surfaces in a geometrical optics approximation is considered, in which two collimated beams of arbitrary irradiance are mapped onto each other. Several methods for phase and irradiance control with double freeform surfaces have been proposed in literature.
\\
One of the first design methods for the mapping of two wavefronts by coupled freeform surfaces is the Simultaneous Multiple Surface (SMS) method, which was developed by Benitez and Mi\~nano \cite{Ben04_1}. The surfaces are thereby constructed from generalized cartesian ovals and by applying constant optical path length (OPL) conditions\cite{chaves2015}. The design method can be utilized for numerous applications in imaging and nonimaging optics\cite{chaves2015, Ben09_1}.
\\
Zhang et al. \cite{Wu14_1} and Shengqian et al. \cite{Wu16_1} solve the design problem by describing it in the form of a Monge-Amp\`ere type PDE, discretizing the equation by finite differences and then solving the resulting nonlinear equation system by the Newton method. The design method can be applied to a variety of wavefront shapes \cite{Wu16_1}.
\\ 
An alternative approach to construct freeform surfaces for irradiance and phase control is from ray mapping methods\cite{Oli13_1, Bau12_1, Brun13_1, Feng13_1, Feng13_2, Feng15_3, Feng16_4, Dosk16_1, Dosk16_2,Boe16_1, Boe16_2, Yadav16_1}. These methods are based on the seperation of the design process into two separate steps: the calculation of an \textit{integrable} ray mapping between the source and the target irradiance and the subsequent construction of the \textit{continuous} freeform surfacesfrom the mapping. The integrability thereby ensures the continuity of the freeform surfaces and the mapping of the input irradiance onto the ouptut irradiance. Since the integrability depends on the physical properties of the optical system it is in general a nontrivial task to find such a mapping.
\\
As shown in several publications, there is a strong relation between the inverse problem of nonimaging optics and optimal mass transport (OMT)\cite{Glimm03_1, Glimm04_2, Wang04_1, Rub07_1, Glimm10_3, Oliker11_2}. The cost function, which has to be applied to a certain optical configuration, is thereby problem-specific. For example, the mapping of two collimated beams with arbitrary irradiance onto each other with double freeform \textit{mirrors} is described by a quadratic cost function \cite{Glimm04_2} and can be solved by corresponding numerical schemes\cite{Yadav16_1}. The same problem statement with double freeform \textit{lenses}, which is considered in this work, is described by a different cost function, which depends on the OPL between the freeform surfaces as it was shown by Rubinstein and Wolansky\cite{Rub07_1}.
\\
The investigations presented here are inspired by several publications \cite{Bau12_1,Brun13_1,Feng13_1,Feng13_2,Feng15_3,Feng16_4}, in which the authors applied the quadratic OMT cost function to calculate a ray mapping to deal with the lens design problem. With this ray mapping, designs have been demonstrated of both single freeform surfaces for irradiance control for collimated input beams and point sources \cite{Bau12_1,Brun13_1,Feng16_4}, and that of double freeform surfaces for irradiance and phase control\cite{Feng13_1,Feng13_2,Feng15_3}. As demonstrated for illumination control with single freeform surfaces in Ref. \cite{Boe16_1} and for collimated beam shaping with double freeform surfaces in Ref. \cite{Boe16_2} the design problems are thereby restricted to a paraxial approximation.
\\
Here we first investigate the design by ray-mapping methods of double freeform surfaces which map two collimated beams with arbitrary irradiance onto each other \textit{beyond} the paraxial approximation. To overcome the restriction to the paraxial regime, which is necessary for the construction of compact systems, the design problem will be modeled by two coupled PDE's. This involves on one hand the Jacobian equation, expressing the local energy conservation, and on the other hand a ray mapping PDE, enforcing the surface continuity and the constant OPL. The PDEs will then be solved by an optimization scheme with the OMT mapping from the quadratic cost function as the intial iterate, leading to a construction approach for the freeform surfaces.
\\
To do so, the work is structured as follows. In section \ref{sec:2}, by using the law of refraction, the constant OPL condition and a surface continuity condition, a PDE for an integrable ray mapping, is derived. Together with the Jacobian equation it builds a system of PDEs for the determination of the mapping components. It is argued that the PDE system is fulfilled within the paraxial approximation by the quadratic cost function OMT map. In section \ref{sec:3}, a method for solving the PDE system for general lens lens distances is presented. It is based on discretizing the PDEs with finite differences and solving the resulting system of nonlinear equations by a standard optimization scheme with the quadratic cost function OMT map as the initial iterate. 
A summary of the design algorithm and a detailed discussion of the implemenation is presented in section \ref{sec:4}, followed by the application of the presented method to the design of a single-lens and a two-lens system in section \ref{sec:5}. Finally, in section \ref{sec:6}, a short discussion of the results is presented.

\section{Freeform design in paraxial approximation}
\label{sec:2} 

\subsection{Energy conservation and cost functions}

In Ref. \cite{Boe16_1} a design method was presented for the construction of a single freeform surface for a collimated input beam with irradiance $I_S (x,y)$ and an arbitrary illumination pattern $I_T (x,y)$ on a target plane. It was shown that in the paraxial approximation the design process can be decoupled into two separate steps. In the first step a raymapping $\mathbf{u}(x,y)=(u_x (x,y), u_y (x,y))$ is calculated from the theory of optimal mass transport, and in the second step the freeform surface is constructed from the mapping.
\\
There are several basic physical principles that a ray mapping needs to fulfill.
Firstly, to map the source irradiance $I_S (x,y)$ onto the target irradiance $I_T (x,y)$, the ray mapping should be energy conserving. The local energy conservation is expressed through the Jacobian Eq.
\begin{gather}\label{eq:1}
det(\nabla \mathbf{u}(x,y))I_T (\mathbf{u}(x,y))=I_S (x,y).
\end{gather}
Secondly, in case of freeform illumination optics, the mapping should allow the calculation of \textit{continuous} freeform surfaces. As shown in several publications, these so called \textit{integrable} ray mappings are related to problem-specific cost functions representing different optical settings, where one has to distinguish between point sources and/or collimated beams, mirrors and/or lenses and so on\cite{Glimm03_1, Glimm04_2, Wang04_1, Rub07_1, Glimm10_3, Oliker11_2}. The cost function defines a metric between the source distribution $I_S (x,y)$ and the target illumination pattern $I_T (x,y)$ and therefore represents an additional constraint to the underdetermined Eq. (\ref{eq:1}). In the case of a single freeform surface for the redistribution of collimated input beams, the quadratic cost function
\begin{equation}\label{eq:2}
d(I_S, I_T)^2 = inf_{\mathbf{u}\in M} \int |\mathbf{u}(\mathbf{x})-\mathbf{x}|^2 I_S (\mathbf{x}) d\mathbf{x},
\end{equation}
which is valid in the paraxial approximation, was studied by the authors\cite{Boe16_1}. A key property there was the vanishing curl
\begin{gather}\label{eq:3}
\partial_y u_x (x,y) -\partial_x u_y (x,y)=0,
\end{gather}
characterizing the quadratic cost function in Eq. (\ref{eq:2}) \cite{Hak04_1}.\\
As shown by Rubinstein and Wolansky, the cost function for collimated beam shaping with double freeform \textit{lenses} takes a different form than Eq. (\ref{eq:2})\cite{Rub07_1}. The authors propose to minimize the corresponding cost function by a steepest descent algorithm to get the ray mapping\cite{Rub07_1}, but unfortunatly a numerically stable implementation is a nontrivial problem.
\\
Due to its applicability in the paraxial approximation (see below) and the availability of numerous published stable numerical schemes for its calculation, the quadratic cost function OMT mapping serves as an initial iterate for the optimization scheme presented below. It will therefore build the basis of the design approach presented in sections \ref{sec:3} and \ref{sec:4}.

\subsection{Ray mapping condition}
\label{sec:2.1} 

\begin{figure}[h]
\begin{center}
\begin{tabular}{c}
\includegraphics[height=6.5cm]{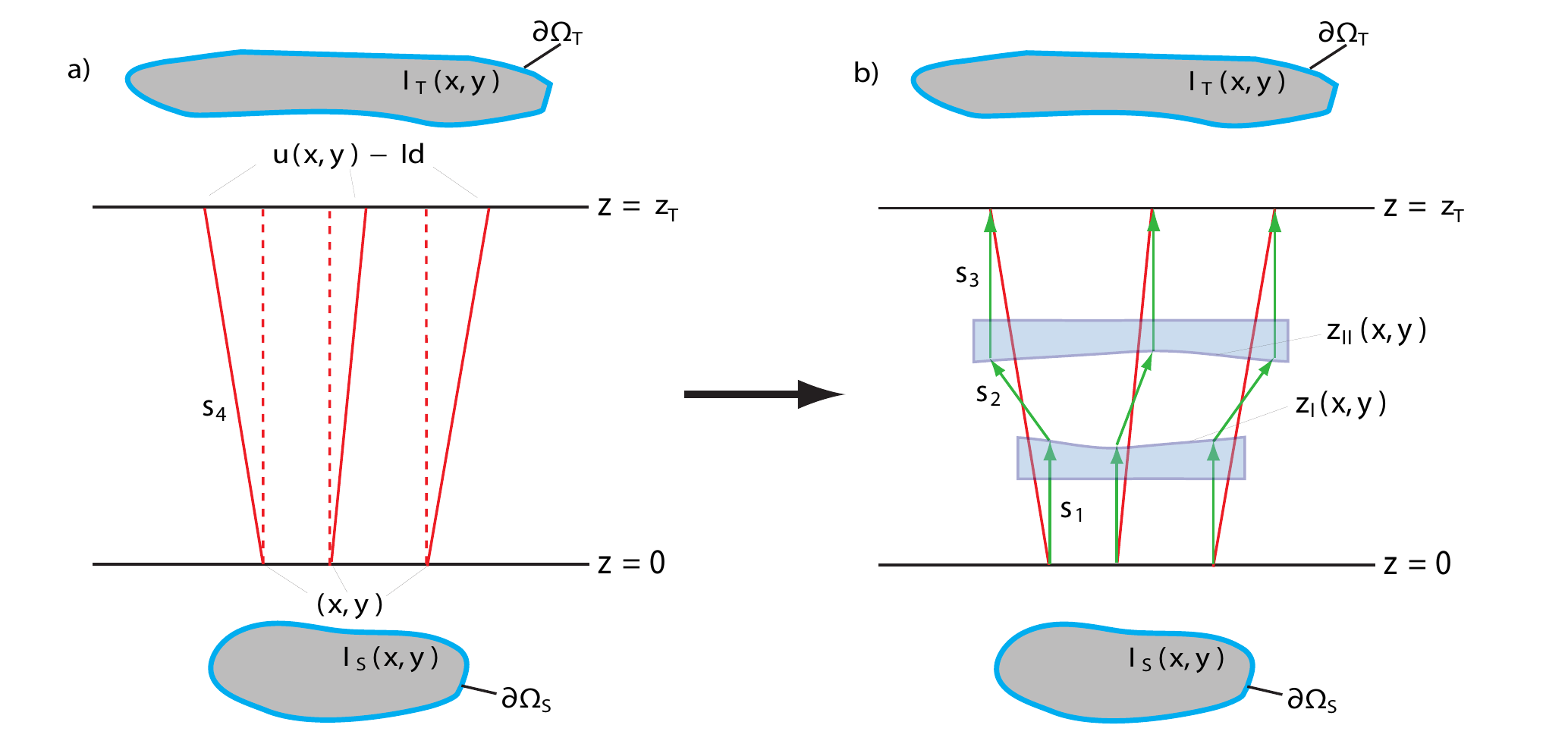}
\end{tabular}
\end{center}
\caption 
{ \label{fig:1}
a) The irradiance distributions $I_S (x,y)$ and $I_T (x,y)$ with the boundaries $\partial \Omega_S$ and $\partial \Omega_T$ are given on the planes $z=0$ and $z=z_T$, respectively.  In the first step an integrable ray mapping $\mathbf{u}(x,y)= (u_x (x,y), u_y (x,y))$ is calculated between the distributions, which defines the vectorfield $\mathbf{s}_4$ between source and target plane. b) In the second step the freeform surfaces $z_{I} (x,y)$ and $z_{II} (x,y)$ are calculated from the law of refraction and the constant OPL condition. The collimated in- and output beams are represented by the vector fields $\mathbf{s}_1$ and $\mathbf{s}_3$ and the refracted beam by the vector field $\mathbf{s}_2$.} 
\end{figure} 

We follow the approach from Ref. \cite{Boe16_1, Boe16_2} by expressing the basic geometry according to Fig. \ref{fig:1} in terms of the collimated input and output vector fields $\mathbf{s}_1$ and $\mathbf{s}_3$, the refracted vector field $\mathbf{s}_2$ and the ray mapping vector field $\mathbf{s}_4$:
\begin{gather}\label{eq:4}
\mathbf{s}_1 =
\begin{pmatrix}
  0 \\
  0 \\
  z_{\textrm{I}} (x,y)
\end{pmatrix}, \ \ \
\mathbf{s}_2 =
\begin{pmatrix}
  u_x (x,y) -x\\
  u_y (x,y) -y\\
  z_{\textrm{II}} (u_x , u_y)-z_{\textrm{I}} (x , y)
\end{pmatrix}, \ \ \\
\mathbf{s}_3 =
\begin{pmatrix}
  0 \\
  0 \\
  z_T - z_{\textrm{II}} (u_x , u_y)
\end{pmatrix}, \ \ \
\mathbf{s}_4=\begin{pmatrix}
  u_x (x,y) -x\\
  u_y (x,y) -y\\
  z_T
\end{pmatrix}. \nonumber
\end{gather}
Since the goal is to calculate at least continuous freeform surfaces, we have to apply the surface continuity condition 
\begin{gather}\label{eq:5}
\mathbf{n}_{\textrm{I}} \cdot (\nabla \times \mathbf{n}_{\textrm{I}})=0
\end{gather}
to both freeform surfaces $z_{\textrm{I}}(x,y)$ and $z_{\textrm{II}}(x,y)$ and their normal vector fields $\mathbf{n}_{\textrm{I}}(x,y)$ and $\mathbf{n}_{\textrm{II}}(x,y)$, respectively.
Thereby the normal vector fields can be expressed with the law of refraction
\begin{gather}\label{eq:6}
\mathbf{n}_{\textrm{I}} = n_1 \hat{\mathbf{s}}_1 -n_2 \hat{\mathbf{s}}_2,
\end{gather}
in terms of the normalized incoming and refracted vector fields $\hat{\mathbf{s}}_1$ and $\hat{\mathbf{s}}_2$ and the refractive indices $n_1$ and $n_2$. Hence Eq (\ref{eq:5}) can be written as
\begin{gather}\label{eq:7}
\mathbf{s}_2(\nabla \times \mathbf{s}_1)
= n_1 \frac{\{\mathbf{s}_2 \times [(\mathbf{s}_2\nabla)\mathbf{s}_2 ]\}_3 }{\mathbf{n}\cdot\mathbf{s}_2}-\mathbf{s}_2(\nabla \times \mathbf{s}_3) +\mathbf{s}_2(\nabla \times \mathbf{s}_4)
\end{gather}
and therefore, by plugging in the vector fields given in Eq. (\ref{eq:4}) and defining  $\mathbf{v}:=(\mathbf{u}(x,y)-\mathbf{Id})^{\perp}=(-(u_y (x,y)-y),u_x (x,y)-x)$, we get
\begin{gather}\label{eq:8}
\mathbf{v}\nabla z_{\textrm{I}}(x,y)
= 
n_1 \cdot \frac{ \mathbf{v}\cdot \left[\left(\mathbf{v}^{\perp}\cdot \nabla \right)\mathbf{v}^{\perp}\right]}{\mathbf{n}\cdot \mathbf{s}_2}
 +
\mathbf{v}\nabla z_{\textrm{II}}(u_x ,u_y )
-(z_{\textrm{II}}(u_x ,u_y )- z_{\textrm{I}}(x,y) )\nabla \mathbf{v}.
\end{gather}
Comparing this equation with the case of a single freeform surface\cite{Boe16_1}, we see that the $\mathbf{v}\nabla z_{\textrm{II}}(u_x ,u_y )$-term on the right hand side (RHS) arises due to the second freeform surface $z_{\textrm{II}}(x,y)$ and is therefore connected to the recollimation of the refracted vector field.
\\
The left hand side (LHS) of Eq. (\ref{eq:8}) represents the dot product of the projected gradient of the first surface $\nabla z_{\textrm{I}}(x,y)$ =$(\partial_x  z_{\textrm{I}}(x,y), \partial_y  z_{\textrm{I}}(x,y))$ and the direction perpendicular to the ray deflection $(\mathbf{u}(x,y)-\mathbf{Id})$. Therefore, a nonvanishing RHS of Eq. (\ref{eq:8}) contradicts the law of refraction, which states that the incoming, the refracted, and the normal vector have to lie in the same plane. This can be seen directly by using the relation 
\begin{gather}\label{eq:9}
\nabla(z- z_{\textrm{I}}(x,y)) \stackrel{!}{=}\frac{\mathbf{n}_{\textrm{I}}(x,y)}{(\mathbf{n}_{\textrm{I}}(x,y) )_z} \ \ \ 
\Leftrightarrow
\ \ \
\begin{pmatrix}
  \partial_x z_{\textrm{I}}(x,y) \\
  \partial_y z_{\textrm{I}}(x,y)
\end{pmatrix}
\stackrel{!}{=}
\begin{pmatrix}
 \frac{n_2 \cdot (u_x -x)}{|\mathbf{s}_2|\cdot(\mathbf{n_{\textrm{I}}})_z } \\
 \frac{n_2 \cdot (u_y -y)}{|\mathbf{s}_2|\cdot(\mathbf{n_{\textrm{I}}})_z }
\end{pmatrix} \propto \mathbf{v}^{\perp},
\end{gather}
leaving us with the condition that the RHS of Eq. (\ref{eq:8}) has to be equal to zero.\\
A further relation can be derived by applying the chain rule to the Eq.
$\nabla_{\mathbf{u}}z_{\textrm{II}}(u_x, u_y) = (\partial_{u_x} z_{\textrm{II}}(u_x, u_y), \partial_{u_x} z_{\textrm{II}}(u_x, u_y))  \stackrel{!}{=}\frac{\mathbf{n_{\textrm{II}}}(u_x , u_y )}{(\mathbf{n_{\textrm{II}}})_z }$\cite{Rub01_1}. This provides us with the gradient $\nabla z_{\textrm{II}}(x,y)$, which is used to rewrite the second term of the RHS of Eq. (\ref{eq:8}).\\
Hence from the continuity condition (\ref{eq:5}) and the law of refraction (\ref{eq:6}) follow the system of Eqs.
\begin{subequations}
\label{eq:10}
\begin{align}
& \mathbf{v}\nabla z_{\textrm{I}}(x,y)
= 0, 
 \label{eq:10a}
 \\
& n_1 \cdot \frac{ \mathbf{v}\cdot \left[\left(\mathbf{v}^{\perp}\cdot \nabla \right)\mathbf{v}^{\perp}\right]}{\mathbf{n}\cdot \mathbf{s}_2}
+
 n_2 \frac{g(u_x , u_y)}{(\mathbf{n_{\textrm{II}}})_z \cdot|\mathbf{s}_2 |}
-
(z_{\textrm{II}}(u_x,u_y)- z_{\textrm{I}}(x,y) )\nabla \mathbf{v}=0
, 
\label{eq:10b}
\end{align}
\end{subequations}
with  $g(u_x , u_y)=-v_x^2 \partial_x u_y +v_y^2 \partial_y u_x +v_x v_y (\partial_x u_x - \partial_y u_y)$ and similar Eqs. by considering Eq. (\ref{eq:5}) and Eq. (\ref{eq:6}) for the second freeform surface $z_{\textrm{II}}(x,y)$.\\
These Eq. (\ref{eq:10}) together with Eq. (\ref{eq:1}) build a system of PDEs for the unknown mapping $\mathbf{u}(x,y)$ and the surfaces $z_{\textrm{I}}(x,y)$ and $z_{\textrm{II}}(x,y)$.\\
To decouple the design process into separate steps as described in the beginning of this section, one needs to find a ray mapping fulfilling the condition (\ref{eq:10b}) which is nontrivial without any a priori knowledge about the freeform surfaces. For single and double freeforms this can be done by considering the small-angle approximation $(z_{\textrm{II}}(u_x,u_y)- z_{\textrm{I}}(x,y))\gg |\mathbf{u}(x,y)-\mathbf{Id}|$ leading to a vanishing first and second term in (\ref{eq:10b}). Addionally, using the mapping from the quadratic cost function defined by Eq. (\ref{eq:3}), the condition (\ref{eq:10b}) is fulfilled and the surfaces can be calculated by using Eq. (\ref{eq:10a}) with appropriate boundary conditions.  The boundary conditions can be derived by considering the law of refraction (\ref{eq:9}) on the boundaries of $I_S (x,y)$ and $I_T (x,y)$ \cite{Boe16_1}. As discussed in Ref. \cite{Boe16_1} , this leads to a path independent integration of Eq. (\ref{eq:9}) to calculate the surface $z_{\textrm{I}}(x,y)$. \\
An alternative way to derive the validity of the quadratic cost function in the paraxial approximation results is by utilizing an expansion of the Rubinstein-Wolansky cost function \cite{Rub07_1} for small angles. \\
The double freeform design process can further be simplified
by also using the constant OPL condition 
\begin{gather}\label{eq:11}
n_1 |\mathbf{s}_1 |+n_2 |\mathbf{s}_2 |+n_1 |\mathbf{s}_3 |\stackrel{!}{=}const \equiv OPL.
\end{gather}
By plugging in Eq. (\ref{eq:4}), the Eq. (\ref{eq:11}) can be solved for
\begin{gather}\label{eq:12}
z_{\textrm{II}}(u_x, u_y)-z_{\textrm{I}}(x, y)=
-\frac{n}{n^2 -1} OPL_{red}\mp \frac{1}{n^2 -1}\sqrt{OPL_{red}^2+(n^2-1)|\mathbf{u}(x,y)-\mathbf{x}|^2}
\end{gather}
with $n:=n_1 /n_2$ and $OPL_{red}:= (OPL-n_1 z_T)/n_2$ between the first and second surface. The sign in Eq. (\ref{eq:12}) depends on whether we have a single-lens ($OPL_{red}>0$; $n<1$: negative sign) or a two-lens system ($OPL_{red}<0$; $n> 1$: positive sign). According to Eq. (\ref{eq:12}), for single-lens systems the mapping values are restricted by the relation $|\mathbf{u}(x,y)-\mathbf{x}|^2 <OPL_{red}^2 /|n^2 -1|$. \\
 Since $\mathbf{n}\cdot \mathbf{s}_2$ and $(\mathbf{n_{\textrm{II}}})_z \cdot|\mathbf{s}_2 |$
depend on $z_{\textrm{II}}(u_x, u_y)-z_{\textrm{I}}(x, y)$, Eq. (\ref{eq:10b}) can be written as a PDE for the components of mapping function $\mathbf{u}(x,y)$ only.
Therefore, the Eqs. (\ref{eq:1}) and (\ref{eq:10b}) build a system of PDEs for the functions $u_x (x,y)$ and $u_y (x,y)$. Both Eqs. build the basis of the design process for double freeform surfaces described in sections \ref{sec:3} and \ref{sec:4}.\\
Before we give any details we want to discuss the condition (\ref{eq:10b}) briefly for freeform mirrors.

\subsubsection{Freeform Mirrors}
\label{sec:2.2}
For mirrors the refractive indices in Eq. (\ref{eq:6}) have to be replaced by $n_{1}\equiv n_{2}\equiv -1$ and we get $\mathbf{n}_{\textrm{I}}\mathbf{s}_2 = -(\mathbf{n}_{\textrm{II}})|\mathbf{s}_2|$.
Therefore, Eq. (\ref{eq:10b}) reduces to
\begin{gather}\label{eq:13}
-\frac{(v_x^2 + v_y^2 )}{\mathbf{n_{\textrm{I}}}\cdot \mathbf{s_2}}
 \nabla \mathbf{v}
 -
(z_{\textrm{II}}(u_x,u_y)- z_{\textrm{I}}(x,y) )\nabla \mathbf{v}\stackrel{!}{=}0,
\end{gather}
Hence the two-refractor problem with collimated beams can be solved if the mapping fulfills $\nabla \mathbf{v}=\partial_y u_x -\partial_x u_y =0$, which is the case for the quadratic cost function defined by Eq. (\ref{eq:1}) and Eq. (\ref{eq:3}). This was proven in a mathematically rigorous way by Glimm and Oliker \cite{Glimm04_2}.\\
Hence, in contrast to the single lens, single mirror and double freeform lens systems, the design problem can be solved by the quadratic cost function without any additional assumptions like the paraxial approximation.

\section{Freeform Lens design beyond paraxial approximation}
\label{sec:3}
As mentioned in the previous section, the Eqs. (\ref{eq:1}) and (\ref{eq:10b}) are the basis of the design approach presented in the following. Since Eq. (\ref{eq:10b}) is exactly fulfilled by the mapping with the quadratic cost functions for an infinite distance between the freeform surfaces or an infinite OPL, respectively, the Eq. (\ref{eq:10b}) represents a correction to Eq. (\ref{eq:3}) for finite OPLs. Hence, for finite distances between $z_{\textrm{I}}(x,y)$ and $z_{\textrm{II}}(x,y)$ we are searching for corrections $\Delta\mathbf{u}(x,y)=(\Delta u_x (x,y), \Delta u_y (x,y))$ with $\Delta\mathbf{u}(x,y)\stackrel{OPL\rightarrow \infty}{\rightarrow} 0$ to the ray mapping defined by the quadratic cost function, which we will denote by $\mathbf{u}^{\infty} (x,y)$ in the following.
Hence, after writing Eq. (\ref{eq:10b}) in terms of Eq. (\ref{eq:3}) plus a perturbation term, we want to solve the system of equations
\begin{subequations}
\label{eq:14}
\begin{align}
&det(\nabla \mathbf{u}(x,y))I_T (\mathbf{u}(x,y))=I_S (x,y), \label{eq:14a}\\
&\partial_y u_x - \partial_x u_y -\frac{n^2 -1}{OPL_{red}^2+(n^2 -1) |\mathbf{u}(x,y)-Id|}\nonumber\\
&\cdot[(u_x -x)^2 \partial_y u_x -(u_y -y)^2 \partial_x u_y +(u_x -x)(u_y -y)(\partial_y u_y - \partial_x u_x)]=0, \label{eq:14b}
\end{align}
\end{subequations}
with
\begin{gather}\label{eq:15}
\mathbf{u}(x,y)=\mathbf{u}^{\infty} (x,y)+\Delta \mathbf{u}(x,y)
\end{gather}
and the given function $\mathbf{u}^{\infty} (x,y)$ for the correction $\Delta \mathbf{u}(x,y)$. The scalability of the mapping correction $\Delta \mathbf{u}(x,y)$ with the parameter $OPL_{red}$ thereby suggests to solve Eq. (\ref{eq:14}) by an optimization method, since $OPL_{red}$ can be reduced step by step with the solution $\Delta \mathbf{u}(x,y)$ from the previous step as an initial iterate. This ensures the convergence of the optimization and can be done until the desired design goal or $OPL_{red}$, respectively, is reached.\\
Additionally, we want to apply boundary conditions to Eq. (\ref{eq:14}). To do so, we use standard boundary conditions for OMT problems by demanding that the boundary of the support of $I_S(x,y)$ is mapped onto the boundary of the support of $I_T(x,y)$. In the case of mapping two unit squares onto each other, like in the example section \ref{sec:5}, we therefore apply
\begin{gather}\label{eq:16}
\Delta u_x(-0.5,y)=\Delta u_x(0.5,y)\stackrel{!}{=}0, \ \ \ y\in [-0.5,0.5],\\
\Delta u_y(x,-0.5)=\Delta u_y(x,0.5)\stackrel{!}{=}0, \ \ \ x\in [-0.5,0.5] \nonumber,
\end{gather}
implying that the edges of the boundary of $I_S (x,y)$ are mapped onto the opposing edges of the boundary of $I_T (x,y)$.\\
We discretize (\ref{eq:14}) using the standard central finite differences for the derivatives of $\Delta u_x (x,y) \rightarrow (\Delta u_x)_{i;j}$ and $\Delta u_y (x,y) \rightarrow (\Delta u_y)_{i;j} $ with $i=1,..., N$; $j=1,...,N$  at the inner points $i=2,..., N-1$; $j=2,...,N-1$ and second-order finite differences at the boundary. For the derivatives of e.g. $\Delta u_x (x,y)$ we get
\begin{gather}\label{eq:17}
\partial_x ( \Delta u_x) 
\rightarrow 
\frac{1}{2\Delta x } [(\Delta u_x)_{i;j+1} - (\Delta u_x)_{i;j-1}],\ \ \ \ \
\partial_y ( \Delta u_x)
\rightarrow
\frac{1}{2\Delta y } [(\Delta u_x)_{i+1;j} - (\Delta u_x)_{i-1;j}]
\end{gather}
for the inner points and 
\begin{gather}\label{eq:18}
\partial_x ( \Delta u_x)
\rightarrow
-\frac{1}{2\Delta x } [3(\Delta u_x)_{i;j+2} - 4(\Delta u_x)_{i;j+1}+(\Delta u_x)_{i;j}],\\
\partial_y ( \Delta u_x)
\rightarrow
-\frac{1}{2\Delta y } [3(\Delta u_x)_{i+2;j} - 4(\Delta u_x)_{i+1;j}+(\Delta u_x)_{i;j}]\nonumber
\end{gather}
on the boundary. A system of $2\cdot N^2$ nonlinear equations for the unknows $(\Delta u_x )_{i;j}$ and $(\Delta u_y )_{i;j}$ is left, which can be solved by standard numerical methods. In the next section, we give an overview of the design algorithm and a detailed discussion of the implementation.

\section{Design algorithm}
\label{sec:4}

The design algorithm for the construction of double freeform surfaces presented in the previous section is summarized in the following. Since some of the steps offer freedom to choose between different applicable numerical techniques, we add some remarks below which are important for the examples in section \ref{sec:5}.

\begin{enumerate}
	\item Calculate the optimal mass transport map with quadratic cost function $\mathbf{u}^{\infty}(x,y)$ between the distributions $I_S (x,y)$ and $I_T (x,y)$.
	\item Discretize the system of Eq. (\ref{eq:14}) with e.g. finite differences (\ref{eq:17}) and (\ref{eq:18}) and apply boundary conditions.
	\item Choose the physical paramters ($n_1, n_2, OPL_{red}$) of the system  and solve the system of nonlinear Eqs. with a predefined tolerance and the inital iterate $(\Delta u_x )_{i;j}=0$ and $(\Delta u_y )_{i;j}=0$, with $i;j=1,...,N$. 
	\item Calculate the freeform surfaces from Eqs. (\ref{eq:9}) and (\ref{eq:12}).
\end{enumerate}

There are numerous publications presenting numerical methods for the calculation of optimal mass transport maps for quadratic cost functions $\mathbf{u}^{\infty}(x,y)$. In the example section below, we choose the numerical scheme developed by Sulman et al.\cite{Sul11_1}. It provides an efficient calculation of $\mathbf{u}^{\infty}(x,y)$, but has some drawbacks like the restriction to square supports of the irradiance distributions $I_S (x,y)$ and $I_T (x,y)$. In \textit{our} implementation of Sulman's algorithm we recognized instabilities if the distributions show large irradiance gradients. For such distributions in the example section \ref{sec:5} we therefore choose a background irradiance $I_T (x,y)>0$ to ensure the convergence.\\
For solving the system of nonlinear Eqs. in step $3$ we have used the \textit{fsolve()} function from MATLABs 2015b optimization toolbox with the trust-region-reflective solver. Providing \textit{fsolve()} with the structure of the Jacobian matrix of the object function allows an efficient calculation of the solution of the nonlinear Eq. system even for a large number of variables. The scalability with the parameter $OPL_{red}$ of the distance of the initial iterate to the solution of Eq. (\ref{eq:14}) suggests that the optimization could be accelerated by using e.g. the Newton algorithm.\\
Solving the Eq. system in the form (\ref{eq:14}), we recognized oscillations in the solution $\Delta\mathbf{u}(\mathbf{x})$. These seem to arise due to the irradiance optimization by the Jacobian Eq. (\ref{eq:14a}), which in contrast to Eq. (\ref{eq:14b}), is already fulfilled to a high degree by the initial iterate $\Delta\mathbf{u}(\mathbf{x})=0$. Hence, we replace the $det(\nabla \mathbf{u}^{\infty}(x,y))I_T (\mathbf{u}(x,y))$-term in (\ref{eq:14a}) by
\begin{gather}\label{eq:19}
det(\nabla \mathbf{u}^{\infty}(x,y))I_T (\mathbf{u}(x,y))\stackrel{(\ref{eq:1})}{\rightarrow} \frac{I_S (x,y)}{I_T (\mathbf{u}^{\infty}(x,y))}I_T (\mathbf{u}(x,y)).
\end{gather}
This replacement contains the wrong assumption that the Jacobian Eq. is exactly fulfilled by $\mathbf{u}^{\infty}(\mathbf{x})$, but in our experience this leads to only minor changes in the local energy preserving property of the mapping by the optimization. The optimized map will therefore solve the Jacobian Eq. to (nearly) the same degree as $\mathbf{u}^{\infty}(\mathbf{x})$, but also Eq. (\ref{eq:14b}) to a high degree. Alternatively, instead of using (\ref{eq:19}), one could also redefine $I_T(x,y)$ with (\ref{eq:1}) and $\mathbf{u}^{\infty}(\mathbf{x})$ to optimize solely for Eq. (\ref{eq:14b}).
\\
To calculate the first freeform we use Eq. (\ref{eq:9}) with Eq. (\ref{eq:12}), which can be integrated along an arbitrary path to give the surface sag. The basic assumption for this path independent integration is the satisfaction of Eq. (\ref{eq:14b}) as mentioned in section \ref{sec:2}. Since Eq. (\ref{eq:14b}) cannot be perfectly fulfilled numerically, the path integration leads to an accumulation of errors and therefore to deviations from the predefined plane wavefront.\\
For the second surface, one could use Eq. (\ref{eq:12}) directly, which gives this surface on a scattered grid. Since we want to test the design algorithm with ray tracing software and due to advantages for possible freeform manufacturing processes, we calculate the second surface on a regular grid by applying the ray tracing formula
\begin{gather}\label{eq:20}
\mathbf{\hat{s}}_2 =\frac{n_1}{n_2}\mathbf{\hat{s}}_1 + \left\{-\frac{n_1}{n_2}\cdot \mathbf{\hat{n}} \cdot \mathbf{\hat{s}}_1 + \sqrt{1-\left(\frac{n_1}{n_2}\right)^2[1-(\mathbf{\hat{n}} \cdot \mathbf{\hat{s}}_1)^2]} \right\}\mathbf{\hat{n}}
\end{gather}
and Eq. (\ref{eq:11}) in the form
\begin{gather}\label{eq:21}
|\mathbf{s}_2 |= OPL_{red}+\frac{n_1}{n_2} [z_{\textrm{II}}(x_m, y_m)-z_{\textrm{I}}(x_s, y_s)].
\end{gather}
Hereby $(x_s, y_s)$ are the unknown initial coordinates of the incoming vector $\mathbf{\hat{s}}_1(x_s, y_s)$ and $(x_m , y_m)$ the predefined target coordinates. By interpolating the surface $z_I(x,y)$, which we will do bilinearly, the Eq. (\ref{eq:20}) gives the normalized refracted vector field $\mathbf{\hat{s}}_2 (x,y)$. To calculate $z_{II}(x,y)$ on the desired grid point $(x_m,y_m)$ we use the refracted vector field $\mathbf{s}_2 (x,y, z_{II}):= |\mathbf{s}_2 |\cdot \mathbf{\hat{s}}_2 (x,y)$  and solve the system
\begin{gather}\label{eq:22}
\mathbf{s}_2 (x_s,y_s, z_{II}(x_m, y_m)) \stackrel{!}{=}(x_m -x_s, y_m -y_s, z_{II}(x_m, y_m)-z_I (x_s ,y_s))
\end{gather}
at every grid point $(x_m,y_m)$ for the unknown values $x_s$, $y_s$ and $z_{II}(x_m, y_m)$.

\section{Examples}
\label{sec:5}

\begin{figure}[h]
\begin{center}
\begin{tabular}{c}
\includegraphics[height=4cm]{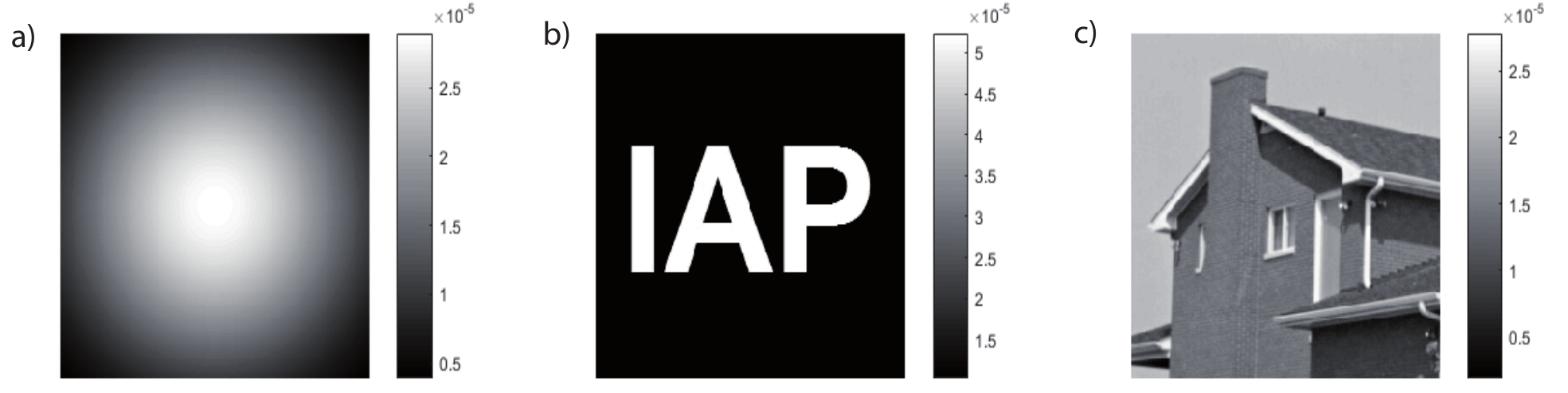}
\end{tabular}
\end{center}
\caption 
{ \label{fig:2}
a) Input irradiance $I_S (x,y)$. b) Output irradiance $I_T (x,y)$ of the example ``IAP". c) Output irradiance $I_T (x,y)$ of the example ``house". The irradiances are normalized to ensure energy conservation.} 
\end{figure}

In the following, the presented design algorithm is applied to two example target distributions. The first example consists of redistributing a gaussian input beam with a waist of $1/\sqrt{2} a.u.$ onto the letters ``IAP", and in the second example the same gaussian will be mapped onto the test image ``house" (see Fig. \ref{fig:2}). In ``IAP" the difficulties arise mainly due to the steep gradients, whereas ``house" shows numerous grey levels. For the resolution of both images $250 \times 250$ pixels are chosen.
\\
Since input and output irradiances are defined on unit squares of the same size, the freeform surfaces have the shape of squares with the side length $1 a.u.$. For the example ``IAP", we choose as the physical parameters of the system, the refractive indices $n_1=1.5$ of the lenses, $n_2=1$ of the surrounding medium and the desired reduced optical path length $OPL_{red}=-0.2 a.u.$ are used. Therefore, a two-lens system like in Fig. \ref{fig:1}b) is considered. An distance between both freeform surfaces of approximately $0.4 a.u.$ can be estimated from Eq. (\ref{eq:12}) since, due to symmetry reasons, the corner points of the irradiances are mapped onto each other.
\\
For the example ``house", we calculate a single-lens system defined by the parameters $n_1=1$, $n_2=1.5$ and  $OPL_{red}=0.2 a.u.$ leading to an approximate distance of $0.6 a.u.$ between the freeform surfaces.\\
We want to point out that it would be possible to design a system with crossing rays for the presented examples. This can be done by mirroring $\mathbf{u}^{\infty}(x,y)$ at the point of origin, which preserves the property (\ref{eq:3}), and using the mirror map for the optimization and according boundary conditions. The initial map $\mathbf{u}^{\infty}(x,y)$ can also be scaled and shifted by constants, which corresponds to a scaling of the size of the target irradiance and shifting of its position, without changing the Eq. (\ref{eq:3}). Additionaly, single-lens systems and two-lens systems can be calculated. Hence, the design method offers the freedom to choose between different optical configurations without a recalculation of the initial map $\mathbf{u}^{\infty}(x,y)$.

\begin{figure}[h]
\begin{center}
\begin{tabular}{c}
\includegraphics[height=12cm]{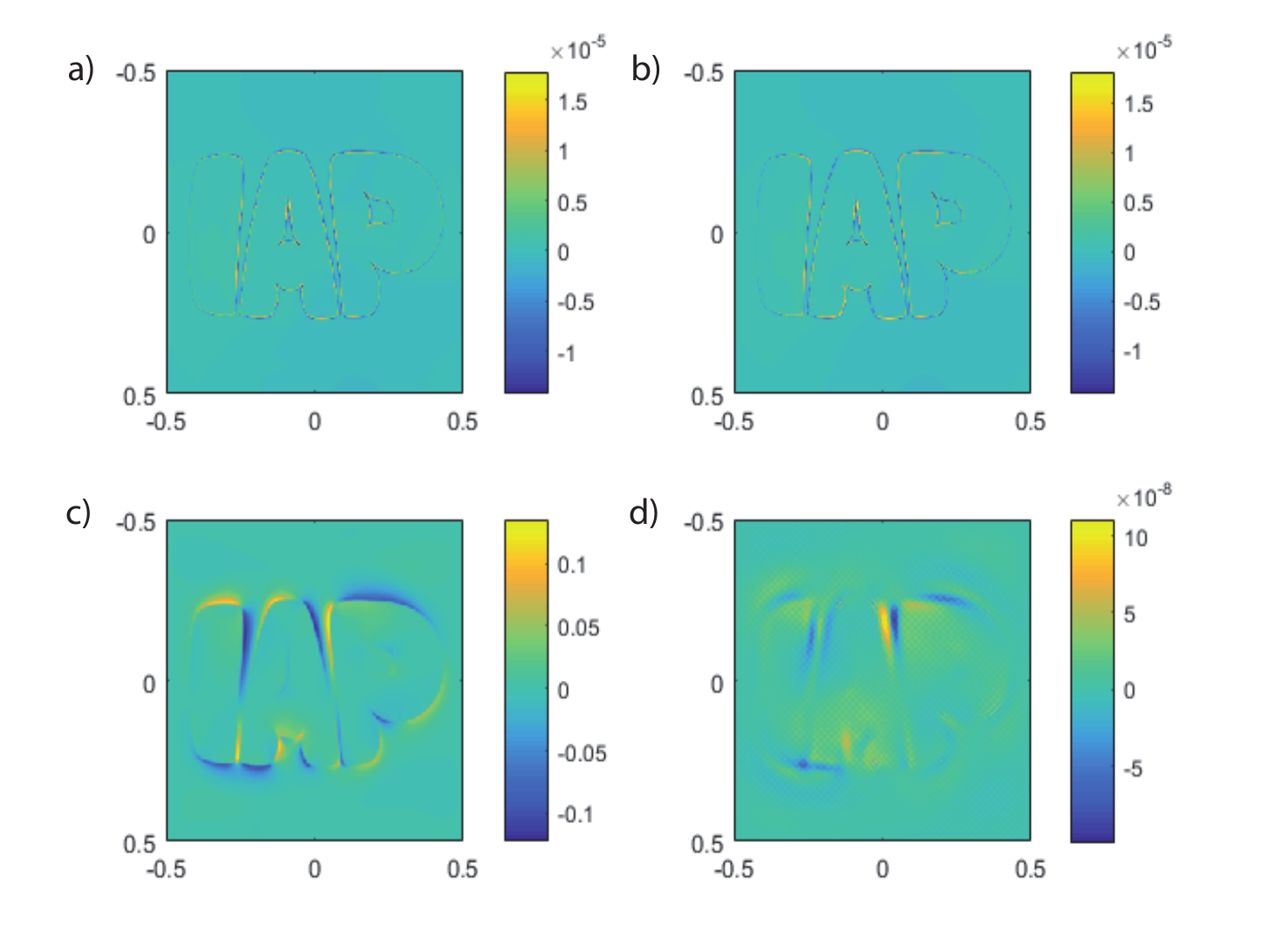}
\end{tabular}
\end{center}
\caption 
{ \label{fig:3}
Jacobian Eq. (\ref{eq:14a}) for ``IAP" a) before the optimzation with $\mathbf{u}^{\infty}(x,y)$ ($rms=2.9629 \cdot 10^{-4}$) and b) after the optimzation with $\mathbf{u}(x,y)$ ($rms=3.1223 \cdot 10^{-4}$). According to the rms values, there is a minimal decrease of quality of the local energy conservation property. Mapping condition (\ref{eq:14b}) c) before the optimization with $\mathbf{u}^{\infty}(x,y)$ ($rms=4.0839$) and b) after the optimization with $\mathbf{u}(x,y)$ ($rms=3.0079\cdot 10^{-6}$). The decrease of the rms leads to an approximate path-independent integration of the map.} 
\end{figure}

\begin{figure}[h]
\begin{center}
\begin{tabular}{c}
\includegraphics[height=12cm]{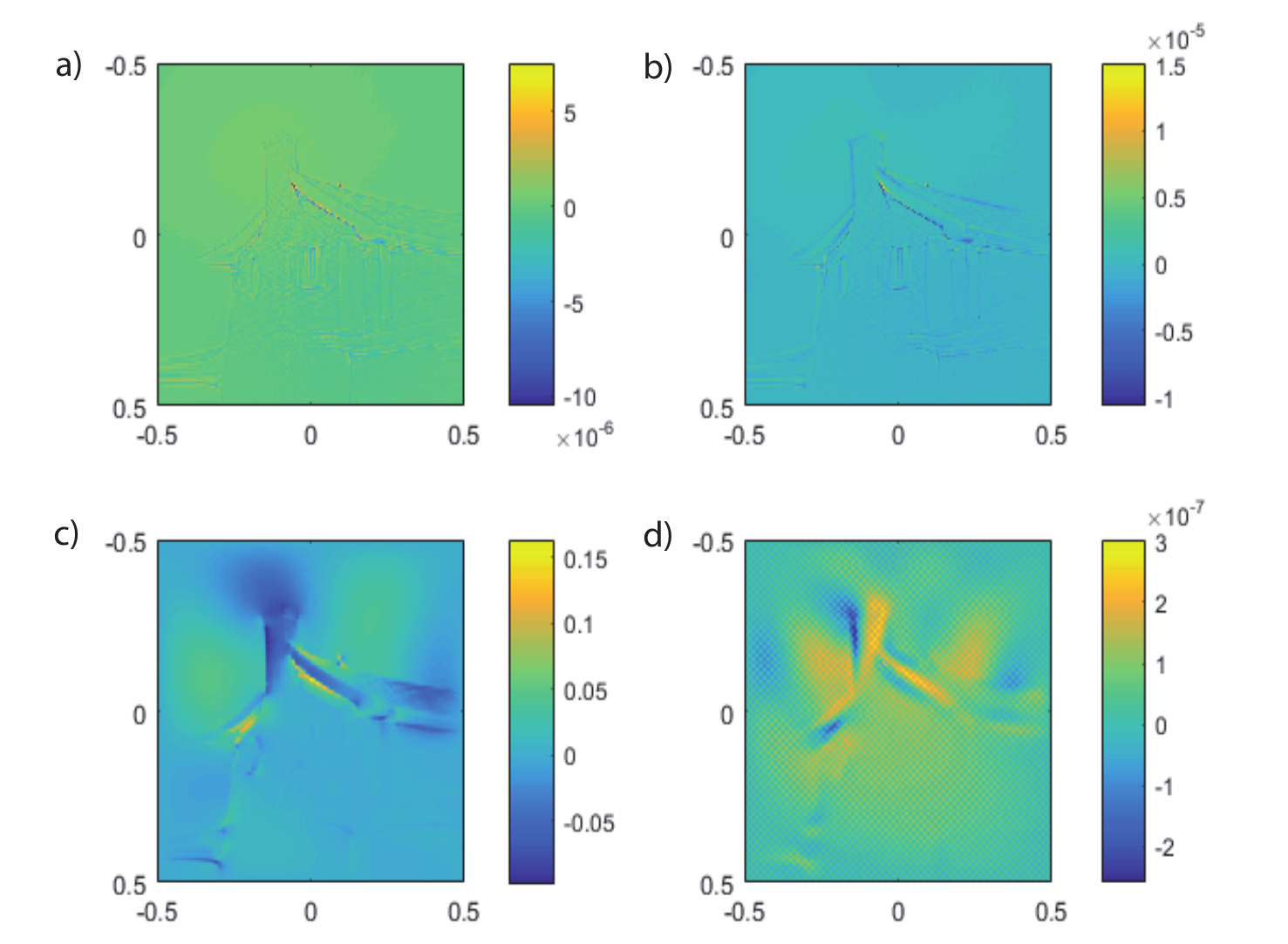}
\end{tabular}
\end{center}
\caption 
{ \label{fig:4}
Jacobian Eq. (\ref{eq:14a}) for ``house" a) before the optimzation with $\mathbf{u}^{\infty}(x,y)$ ($rms=1.4167 \cdot 10^{-4}$) and b) after the optimzation with $\mathbf{u}(x,y)$ ($rms=1.6807 \cdot 10^{-4}$). Again, there is a minimal decrease of quality of the local energy conservation property. Mapping condition (\ref{eq:14b})  c) before the optimization with $\mathbf{u}^{\infty}(x,y)$ ($rms=5.0956$) and b) after the optimization with $\mathbf{u}(x,y)$ ($rms=1.8172\cdot 10^{-5}$).} 
\end{figure}

The calculation of the initial map $\mathbf{u}^{\infty}(x,y)$ with our implementation of Sulmans algorithm took $188 sec$ for ``IAP" and $575 sec$ for ``house" on an Intel Core i3 at $2\times2.4$Ghz with $16$GB RAM.\\
After fixing the physical properties, the mapping has to be optimized according to sections \ref{sec:3} and \ref{sec:4} to obtain the solution of the Eq. system (\ref{eq:14}). To do that, one could use a starting value smaller/larger than the design goal $OPL_{red}=-0.2 a.u.$ or $OPL_{red}=0.2 a.u.$, respectively, to ensure the convergence to the solution of (\ref{eq:14}). However, in practice we did not experience any significant benefit from using smaller/larger starting values for $OPL_{red}$ than the design goal.
For the tolerance of MATLABs \textit{fsolve()}, the value of $10^{-6}$ was used. For both examples the optimization processes with $125,000$ design variables took about $67sec$ for ``IAP" and $71sec$ for ``house''.\\
In Fig. \ref{fig:3} (``IAP") and Fig. \ref{fig:4} (``house") the Eqs. (\ref{eq:14}) are plotted with the initial map $\mathbf{u}^{\infty}(x,y)$ and the optimized map $\mathbf{u}(x,y)$, respectively. Whereas the solutions of the Jacobian Eq. (\ref{eq:14a}), according to the calculated rms values, remains nearly the same, the solution  of the mapping Eq. (\ref{eq:14b}) improves drastically in both cases.

\begin{figure}[h]
\begin{center}
\begin{tabular}{c}
\includegraphics[height=6.5cm]{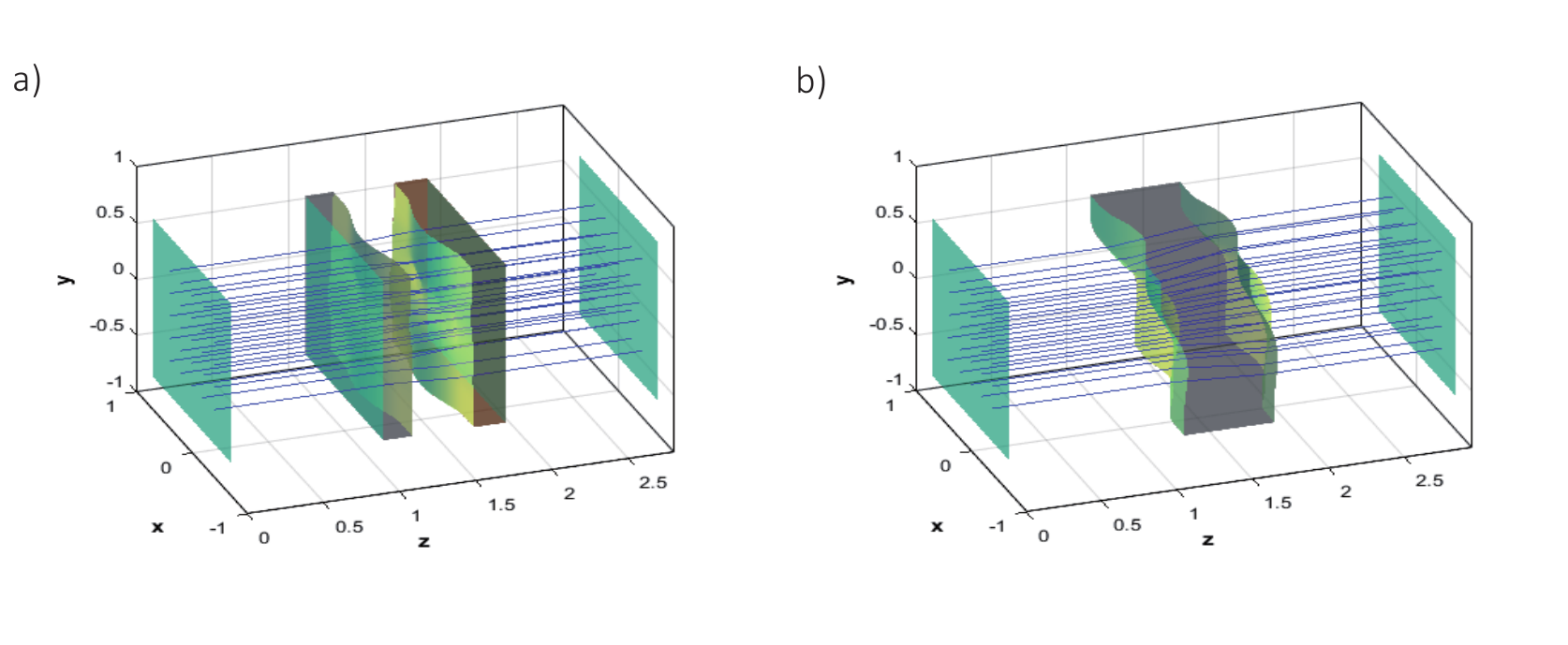}
\end{tabular}
\end{center}
\caption 
{ \label{fig:5}
 Layout of the lens system with freeform surfaces $z_I (x,y)$ and $z_{II} (x,y)$ mapping a collimated gaussian beam onto a collimated beam a) with the ``IAP" irradiance distribution by a two-lens system and b) with the ``house" target irradiance distribution by a single-lens system.} 
\end{figure}

The construction of the surfaces is done by the integration of Eqs. (\ref{eq:9}) and (\ref{eq:12}), first from $(x,y)=(-0.5,-0.5)$ along the x-direction and then along the y-direction. The system layouts with the freeform surfaces for both examples together with a few rays can be seen in Fig. \ref{fig:5}. 

\begin{figure}[h]
\begin{center}
\begin{tabular}{c}
\includegraphics[height=16cm]{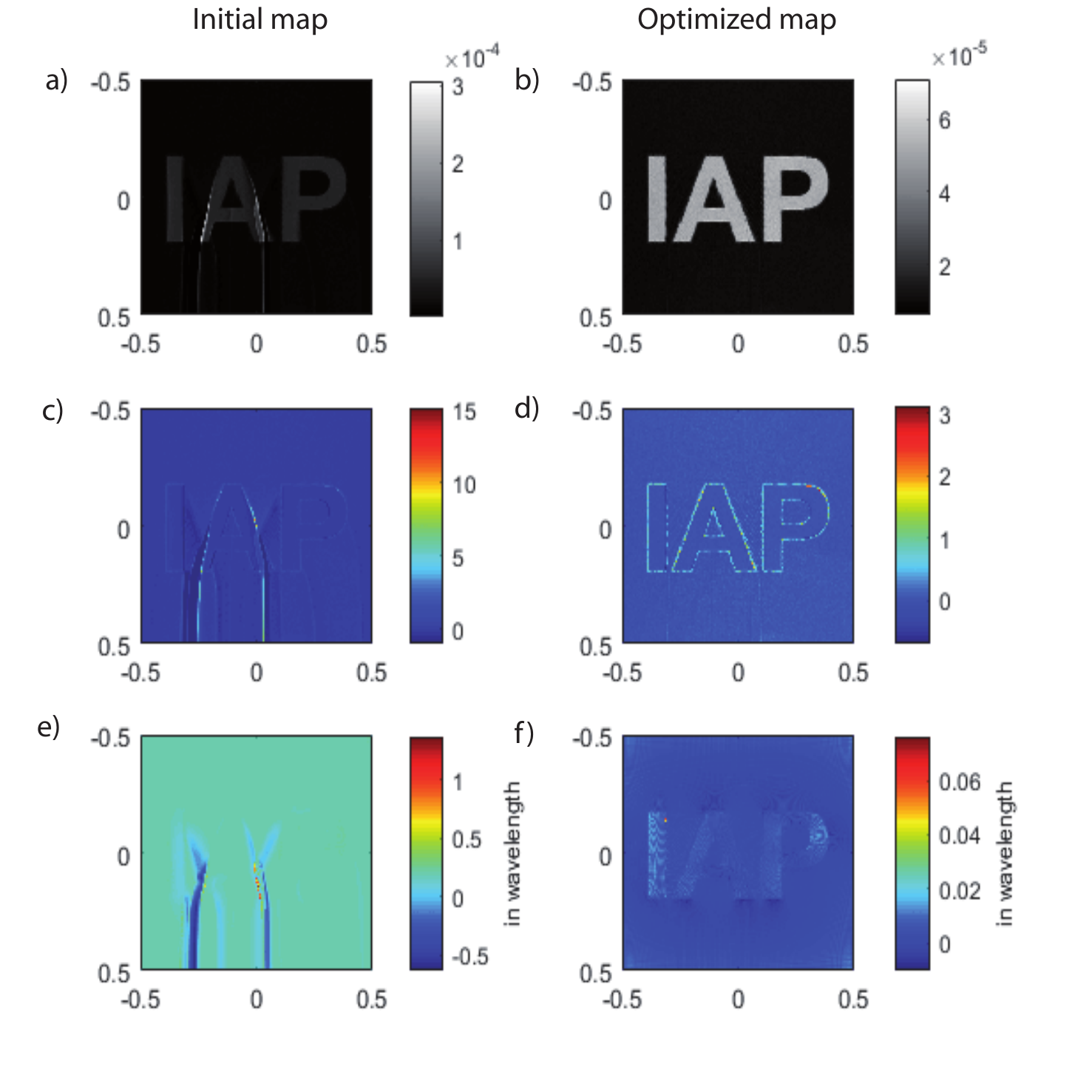}
\end{tabular}
\end{center}
\caption 
{ \label{fig:6}
Results from the ray tracing evaluation for the predefined irradiance distribution ``IAP". Output irradiance distribution from the ray tracing using surfaces from a) the initial map $\mathbf{u}^{\infty}(x,y)$  and b) the optimized map $\mathbf{u}(x,y)$. Normalized difference $\Delta I_T(x,y)$ between predefined (Fig. \ref{eq:2}b) and irradiance distribution from the ray tracing using surfaces from c) the initial map $\mathbf{u}^{\infty}(x,y)$ ($rms=0.43346$) and d) the optimized map $\mathbf{u}(x,y)$ ($rms=0.15183$). Optical path difference from the ray tracing with a reference wavelength of $\lambda=550nm$ using surfaces from e) the initial map $\mathbf{u}^{\infty}(x,y)$ ($rms=0.19515 \lambda$) and f) the optimized map $\mathbf{u}(x,y)$ ($rms=0.00420 \lambda $). Since the maps are integrated first from $(-0.5,-0.5)$ along the $x$-direction and then along the $y$-direction, the deviations from the plane wavefront in e) are consistent with Fig. \ref{fig:3}c).} 
\end{figure}

\begin{figure}[h]
\begin{center}
\begin{tabular}{c}
\includegraphics[height=16cm]{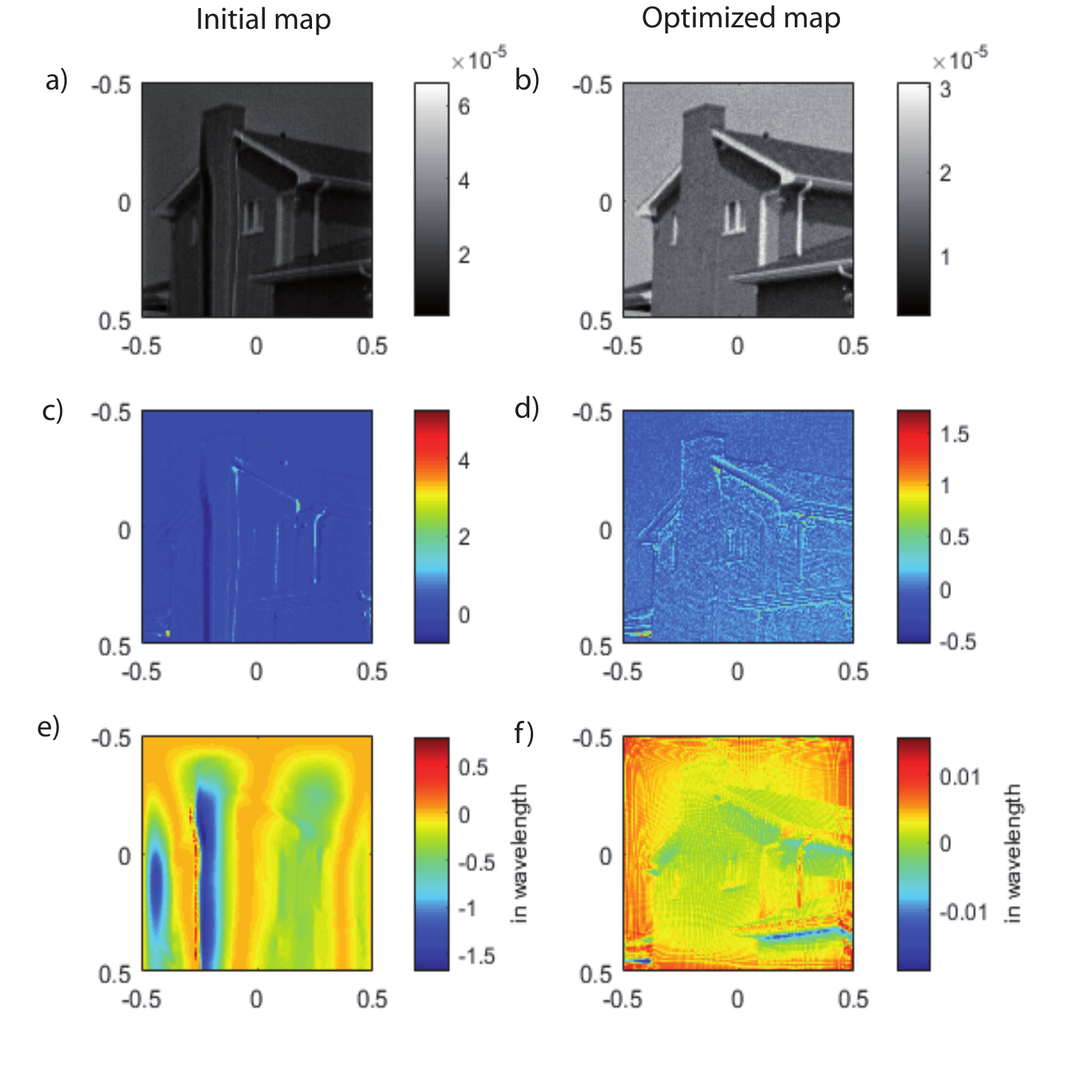}
\end{tabular}
\end{center}
\caption 
{ \label{fig:7}
Results from the ray tracing evaluation for the predefined irradiance distribution ``house". Output irradiance distribution from the ray tracing using surfaces from a) the initial map $\mathbf{u}^{\infty}(x,y)$  and b) the optimized map $\mathbf{u}(x,y)$. Normalized difference $\Delta I_T(x,y)$ between predefined (Fig. \ref{eq:2}c) and irradiance distribution from the ray tracing with a reference wavelength of $\lambda=550nm$ using surfaces from c) the initial map $\mathbf{u}^{\infty}(x,y)$ ($rms=0.22283$) and d) the optimized map $\mathbf{u}(x,y)$ ($rms=0.08880$). Optical path difference from the ray tracing using surfaces from e) the initial map $\mathbf{u}^{\infty}(x,y)$ ($rms=0.25576\lambda$) and f) the optimized map $\mathbf{u}(x,y)$ ($rms=0.00395\lambda$).} 
\end{figure}

To evaluate the improvement by the optimized map, the freeform surfaces are calculated from the initial mapping $\mathbf{u}^{\infty}(x,y)$ and the optimized map $\mathbf{u}(x,y)$ and imported into ray tracing software for the calculation and the comparison of the illumination patterns and the wavefronts. The results  of the ray tracing with $200\cdot 10^6$ rays can be seen in Fig. \ref{fig:6} (``IAP") and Fig. \ref{fig:7} (``house"). The rms values of the normalized difference $\Delta I_T(x,y):=(I_{T}(x,y)-I_{T,RT}(x,y))/I_{T}(x,y)$ between predefined distribution $I_{T}(x,y)$ from Fig. \ref{fig:2} and the ray tracing illumination patterns $I_{T,RT}(x,y)$ as well as the rms values of the optical path difference show a significant improvement of quality of the illumination patterns and the wavefronts for both examples.\\

The freeform calculation with the initial maps $\mathbf{u}^{\infty}(x,y)$  shows strong deviations from the predefined specifications, whereas the optimized maps $\mathbf{u}(x,y)$ show high quality illumination patterns and a wavefront uniformity beyond the diffraction limit. Remaining deviations from the ideal wavefront result (besides fundamental numerical limitiations) are from Eq. (\ref{eq:14b}), which is not fulfilled exactly and therefore leads to an error accumulation along the integration path of Eqs. (\ref{eq:9}) and (\ref{eq:12}). The precision of the illumination pattern on the other hand is mainly limited by the precision of $\mathbf{u}^{\infty}(x,y)$. The main deviations from the predefined distribution $I_T (x,y)$ are resulting from steep gradients, which can be seen especially for the example ``IAP" in Fig. \ref{fig:6}d). This is in agreement with Fig. \ref{fig:3}b) and observations which were made in Ref. \cite{Boe16_1} .

\section{Conclusion}
\label{sec:6}

A design method for the calculation of compact continuous double freeform surfaces for collimated beam shaping with complex irradiance patterns was presented. The method is based on the ray mapping condition (\ref{eq:14b}), which was derived from the law of refraction and the surface continuity condition in section \ref{sec:2} and builds together with the Jacobian Eq. (\ref{eq:14a}) a system of nonlinear PDEs for the unknown ray mapping $\mathbf{u} (x,y)$.\\
Due to the satisfaction of Eq. (\ref{eq:14}) for infinite lens distances by the mapping from OMT with the quadratic cost function (\ref{eq:2}), this mapping serves as an ideal initial iterate for an optimization approach for solving the system of Eq. (\ref{eq:14}). As it was shown by approximating the Eqs. (\ref{eq:14}) by finite differences and using a standard optimization scheme from MATLAB's optimization toolbox one can ensure a fast convergence to the solution of the Eqs. (\ref{eq:14}). This was demonstrated by applying the presented method to two design examples with complex target distributions and validating the results by ray tracing. The double freeform surfaces showed thereby high accuracy for the irradiance patterns and the wavefront, which was assessed by the calculation of the corresponding $rms$ values of the normalized differences.\\
Further improvements can be made by using OMT methods for more complex boundaries of the source and target distributions, which requires the replacement of Sulman's method (with e.g. \cite{ben14_1}) and the generalization of Eq. (\ref{eq:16}) to more complex boundary shapes. The scalability of the distance of the initial map $\mathbf{u}^{\infty}(x,y)$ by $OPL_{red}$ to the solution of the Eqs. (\ref{eq:14}) suggests the application of e.g. the Newton algorithm for a faster optimization.\\
In our future research, we want to generalize the presented approach to double freeform surfaces for wavefronts different from the plane case, like e.g. spherical wavefronts.

\acknowledgments 
The authors want to thank Ralf Hambach and Mateusz Oleszko for valuable discussions, Johannes Stock for comments on the manuscript and David Musick for a grammar and spelling check. We also want to acknowledge the Federal Ministry of Education and Research Germany for financial support through the project fo+ (FKZ: O3WKCK1D) and KoSimO (FKZ:031PT609X).

%



\vspace{1ex}

\end{spacing}
\end{document}